
\magnification=\magstep1
\hsize=13cm

\def\m@th{\mathsurround=0pt}

\def\fsquare(#1,#2){
\hbox{\vrule$\hskip-0.4pt\vcenter to #1{\normalbaselines\m@th
\hrule\vfil\hbox to #1{\hfill$\scriptstyle #2$\hfill}\vfil\hrule}$\hskip-0.4pt
\vrule}}

\def\addsquare(#1,#2){\hbox{$
	\dimen1=#1 \advance\dimen1 by -0.8pt
	\vcenter to #1{\hrule height0.4pt depth0.0pt%
	\hbox to #1{%
	\vbox to \dimen1{\vss%
	\hbox to \dimen1{\hss$\scriptstyle~#2~$\hss}%
	\vss}%
	\vrule width0.4pt}%
	\hrule height0.4pt depth0.0pt}$}}

\def\Addsquare(#1,#2){\hbox{$
	\dimen1=#1 \advance\dimen1 by -0.8pt
	\vcenter to #1{\hrule height0.4pt depth0.0pt%
	\hbox to #1{%
	\vbox to \dimen1{\vss%
	\hbox to \dimen1{\hss$~#2~$\hss}%
	\vss}%
	\vrule width0.4pt}%
	\hrule height0.4pt depth0.0pt}$}}

\def\Fsquare(#1,#2){
\hbox{\vrule$\hskip-0.4pt\vcenter to #1{\normalbaselines\m@th
\hrule\vfil\hbox to #1{\hfill$#2$\hfill}\vfil\hrule}$\hskip-0.4pt
\vrule}}

\def\naga{%
	\hbox{$\vcenter to 0.4cm{\normalbaselines\m@th
	\hrule\vfil\hbox to 1.2cm{\hfill$\cdots$\hfill}\vfil\hrule}$}}

\def\Flect(#1,#2,#3){
\hbox{\vrule$\hskip-0.4pt\vcenter to #1{\normalbaselines\m@th
\hrule\vfil\hbox to #2{\hfill$#3$\hfill}\vfil\hrule}$\hskip-0.4pt
\vrule}}

\def\PFlect(#1,#2,#3){
\hbox{$\hskip-0.4pt\vcenter to #1{\normalbaselines\m@th
\vfil\hbox to #2{\hfill$#3$\hfill}\vfil}$\hskip-0.4pt}}

\dimen1=0.5cm\advance\dimen1 by -0.8pt

\def\vnaka{\normalbaselines\m@th\baselineskip0pt\offinterlineskip%
	\vrule\vbox to 0.6cm{\vskip0.5pt\hbox to
\dimen1{$\hfil\vdots\hfil$}\vfil}\vrule}

\dimen2=1.5cm\advance\dimen2 by -0.8pt

\def\vnakal{\normalbaselines\m@th\baselineskip0pt\offinterlineskip%
	\vrule\vbox to 1.2cm{\vskip7pt\hbox to
\dimen2{$\hfil\vdots\hfil$}\vfil}\vrule}

\noindent
\vskip2.5cm
\centerline{{\bf Functional Relations and Analytic Bethe Ansatz}}\par
\centerline{\bf for}
\centerline{{\bf Twisted Quantum Affine Algebras}\footnote*{Short title:
Twisted quantum affine algebras}} \vskip2.0cm
\centerline{Atsuo Kuniba\footnote\dag{
E-mail: atsuo@hep1.c.u-tokyo.ac.jp}}
\centerline{Institute of Physics, University of Tokyo}
\centerline{Komaba 3-8-1, Meguro-ku, Tokyo 153 Japan}
\centerline{ and }
\centerline{Department of Mathematical Science, Kyushu University}
\centerline{Fukuoka 812, Japan}
\vskip0.5cm
\centerline{Junji Suzuki\footnote\ddag{
E-mail: jsuzuki@tansei.cc.u-tokyo.ac.jp}}
\centerline{Institute of Physics, University of Tokyo}
\centerline{Komaba 3-8-1, Meguro-ku, Tokyo 153 Japan}
\vskip0.3cm
\vskip5.0cm
\centerline{\bf Abstract}
\vskip0.2cm
\par
Functional relations are proposed for
transfer matrices of solvable vertex models
associated with the twisted quantum affine algebras
$U_q(X^{(\kappa)}_n)$ where
$X^{(\kappa)}_n = A^{(2)}_n, D^{(2)}_n, E^{(2)}_6$ and
$D^{(3)}_4$.
Their solutions are obtained for $A^{(2)}_n$ and conjectured
for $D^{(3)}_4$ in the dressed vacuum form
in the analytic Bethe ansatz.
\vfill
\eject
\noindent
{\bf 1. Introduction}\par
Solvable lattice models in two-dimensions have
a commuting family of row-to-row transfer matrices [1].
In [2], a set of functional relations, the {\it $T$-system},
are proposed for the vertex and RSOS models
associated with any non-twisted quantum affine algebra
$U_q(X^{(1)}_r)$.
In the QISM terminology [3], the $T$-system connects the
transfer matrices with various fusion types in the auxiliary
space but acting on a common quantum space.
It generalizes earlier functional relations [4-8]
and enables the calculation of various physical quantities [9].
In particular, one can derive eigenvalue
formulas of the transfer matrices for fusion
vertex models
in the dressed vacuum form
(DVF) by combining the $T$-system with
the analytic Bethe ansatz.
This program has been executed extensively in [10-12],
where curious Yangian analogues of the
Young tableaux have emerged as natural
objects describing the DVFs.
\par
The aim of this paper is to extend these
recent works to the case of the twisted quantum affine algebras
$U_q(X^{(\kappa)}_n)$ where
$X^{(\kappa)}_n = A^{(2)}_n, D^{(2)}_n, E^{(2)}_6$ and
$D^{(3)}_4$.
After some preliminaries in the next section,
we propose the $T$-system for solvable vertex models
associated with quantum $R$-matrices of $U_q(X^{(\kappa)}_n)$
in section 3.
The structure of the $T$-system is closely
analogous to the realization $(X_n,\sigma)$ of
$X^{(\kappa)}_n$ via the classical Lie algebra
$X_r$ and the Dynkin diagram automorphism
$\sigma$ of order $\kappa$.
Actually it has the same form as that for the non-twisted
case $X^{(1)}_n$ expect the ``modulo $\sigma$ relations".
See (3.4).
Owing to the commutatibity (see (3.2)),
one may regard the $T$-system as functional relations
on the eigenvalues and seek solutions in the DVFs
from the analytic Bethe ansatz viewpoint.
This is the subject of the subsequent sections.
In section 4, we indicate how the
solutions can be constructed from those
of the non-twisted case $X^{(1)}_n$ in general.
As an example we present the full answer
for $A^{(2)}_n$ $T$-system in section 5.
In section 6, we give a conjectural solution to
the $D^{(3)}_4$ $T$-system.
With a slight modification
it also yields a full conjecture on $D^{(1)}_4$ case.
These DVFs are neatly described by Yangian analogues
of the Young tableaux as in [11,12].
Section 7 is devoted to summary and discussion.
\par\vskip0.5cm\noindent
{\bf 2. Twisted quantum affine algebras}\par
Let
$X^{(\kappa)}_n$ be one of the
twisted affine Lie algebras
$A^{(2)}_n\, (n \ge 2),
D^{(2)}_n\, (n \ge 4), E^{(2)}_6$ and
$D^{(3)}_4$, which are realized with the pair
$(X_n, \sigma)$ of the classical simple Lie algebra $X_n$
and the Dynkin diagram automorphism $\sigma$ of order
$\kappa=1,2$ or $3$ [13].
We write
the set of the nodes on the diagram as
${\cal S} = \{1,2, \ldots, n\}$
(numeration as in [2]).
Then $\sigma: {\cal S} \rightarrow {\cal S}$ is the
following map.
$$\eqalign{
A^{(2)}_n:& \quad\sigma(a) = n+1-a, \, a \in \{1,2 \ldots, n\},\cr
D^{(2)}_n:& \quad\sigma(a) =
\cases{a& $a \in \{1,2 \ldots, n-2\}$\cr
       n-1 & $a=n$\cr
       n & $a=n-1$\cr},\cr
E^{(2)}_6:& \quad\sigma(a) =
\cases{6-a&$a \in \{1,2,3,4,5\}$\cr
       6&$a=6$\cr},\cr
D^{(3)}_4:& \quad\sigma(1) = 3,\, \sigma(2) = 2,\,
\sigma(3) = 4,\, \sigma(4) = 1.\cr}
\eqno(2.1)
$$
Let $\hat{\cal S} =\{1,2, \ldots, r\}$
denote the set ${\cal S}$
divided by the identification via $\sigma$.
We explicitly specify the identification map
$\wedge : {\cal S} \rightarrow \hat{\cal S}$ as follows.
$$\eqalign{
X^{(2)}_n = A^{(2)}_{2r}, A^{(2)}_{2r-1}
:&\quad  \hat{a} = \hbox{min}(a,n+1-a),\cr
X^{(2)}_n = D^{(2)}_{r+1} :&\quad \hat{a} =
\cases{a& $a \in \{1,2 \ldots, r-1\}$\cr
       r & $a=r, r+1$\cr},\cr
X^{(2)}_n = E^{(2)}_6 (r = 4) :&\quad \hat{a} =
\cases{\hbox{min}(a,6-a)&$a \neq 6$\cr
       4&$a=6$\cr},\cr
X^{(2)}_n = D^{(3)}_4 (r = 2) :&\quad \hat{1} = 1,\, \hat{2} = 2,\,
\hat{3} = 1,\, \hat{4} = 1.\cr}
\eqno(2.2{\rm a})
$$
We shall keep this relation between $n$ and $r$ throughout.
We also introduce the natural embedding
$\cdot : \hat{\cal S} \rightarrow {\cal S}$ by
$$
\dot{a} = \cases{6 & if $X^{(\kappa)}_n = E^{(2)}_6$ and $a = 4$\cr
                  a & otherwise\cr}.
\eqno(2.2{\rm b})
$$
Obviously, the composition
$
\hat{\cal S} \buildrel \cdot \over \rightarrow
{\cal S} \buildrel \sigma \over \rightarrow
{\cal S} \buildrel \wedge \over \rightarrow
\hat{\cal S}
$
is the identity and so is the
restriction of
${\cal S} \buildrel \wedge \over \rightarrow
\hat{\cal S} \buildrel \cdot \over \rightarrow
{\cal S}$ on the image of
$\hat{\cal S} \buildrel \cdot \over \rightarrow
{\cal S}$.
Consider the quantized universal enveloping
algebra $U_q(X^{(\kappa)}_n)$ of $X^{(\kappa)}_n$
introduced in [14].
In this paper we assume that
the deformation parameter $q$ is generic and put
$$q = e^{\hbar},$$
where $\hbar$ is a parameter.
Let $R_{W,W^\prime}(u)$ be the quantum $R$-matrix [14]
that intertwines the tensor products
$W\otimes W^\prime$ and $W^\prime\otimes W$ of the
finite dimensional irreducible $U_q(X^{(\kappa)}_n)$-modules
$W$ and $W^\prime$.
It obeys the Yang-Baxter equation [1]
$$R_{W,W^\prime}(u)R_{W,W^{\prime\prime}}(u+v)
R_{W^\prime,W^{\prime\prime}}(v)
= R_{W^\prime,W^{\prime\prime}}(v)R_{W,W^{\prime\prime}}(u+v)
R_{W,W^\prime}(u),\eqno(2.3)$$
where $u, v \in {\bf C}$ are the spectral parameters.
$R_{W,W^\prime}(u)$ is a rational function of $q^u$,
which can be properly normalized to be pole-free.
\par
In this paper we shall exclusively consider the
family of the irreducible finite dimensional $U_q(X^{(\kappa)}_n)$-modules
$\{ \hat{W}^{(a)}_m \mid a \in \hat{\cal S}, m \in {\bf Z}_{\ge 1} \}$.
If $X^{(\kappa)}_n = A^{(2)}_n$ or $D^{(2)}_n$,
the module $\hat{W}^{(1)}_1$ is the affinization
of the vector representation
for which the associated $R$-matrix
$R_{\hat{W}^{(1)}_1,\hat{W}^{(1)}_1}(u)$ has been calculated
in [15].
See also the appendix in [16].
For these algebras, $R_{W,W^\prime}(u)$
for the whole family $W, W^\prime \in \{\hat{W}^{(a)}_m \}$
will be obtained by the fusion [17] of the above $R$-matrix,
except those associated with the spinor-like series
$\hat{W}^{(r)}_m$ in $D^{(2)}_{r+1}$.
In general, $\hat{W}^{(a)}_m$ is an analogue of the $m$-fold
symmetric tensor representation of
$\hat{W}^{(a)}_1$.
If one denotes by $W^{(a)}_m$ the
irreducible finite dimensional $U_q(X^{(1)}_n)$-module sketched in [2],
one has
$$
\dim \hat{W}^{(a)}_m = \dim W^{(\dot{a})}_m,\quad
a \in \hat{\cal S}. \eqno(2.4)
$$
The RHS here is essentially the
quantity $Q^{(\dot{a})}_m$ in [2,18] and computable as a certain sum of
the dimensions of irreducible $X_n$-modules [2,18,19].
See also (3.10).
Although these descriptions of $\hat{W}^{(a)}_m$
are conjectural and incomplete in general, they are
very consistent with the loop algebra realization of
$X^{(\kappa)}_n$ [13] as well as
the analytic Bethe ansatz studies in [20]
and the sections 4, 5 and 6.
\par\vskip0.5cm\noindent
{\bf 3. $T$-system for twisted quantum affine algebras}\par
Now we turn to the transfer matrix
$$T^{(a)}_m(u) = \hbox{Tr}_{\hat{W}^{(a)}_m}\Bigl(
R_{\hat{W}^{(a)}_m,\hat{W}^{(p)}_s}(u-w_1) \cdots
R_{\hat{W}^{(a)}_m,\hat{W}^{(p)}_s}(u-w_N)\Bigr). \eqno(3.1)$$
Here $N$ denotes the system size, $w_1, \ldots w_N$ are
complex parameters representing the inhomogeneity,
$p \in \hat{\cal S}$ and $s \in {\bf Z}_{\ge 1}$.
The trace (3.1) is the row-to-row transfer matrix with
the auxiliary space
$\hat{W}^{(a)}_m$ acting on the quantum space
$(\hat{W}^{(p)}_s)^{\otimes N}$.
(More precisely, $\hat{W}^{(a)}_m(u)$ and
$\otimes_{j=1}^N \hat{W}^{(p)}_s(w_j)$, respectively.)
We have suppressed the
quantum space dependence on the LHS of (3.1)
reserving the letters $p$ and
$s$ for this meaning throughout.
Thanks to the Yang-Baxter equation (2.3),
the transfer matrices (3.1) form a
commuting family
$$[T^{(a)}_m(u), T^{(a^\prime)}_{m^\prime}(u^\prime)] = 0.\eqno(3.2)$$
We shall write the eigenvalues of $T^{(a)}_m(u)$ as
$\Lambda^{(a)}_m(u)$.
Our purpose here is to propose
the $T$-system, a set of functional relations, among
the transfer matrices
$\{T^{(a)}_m(u) \mid a \in \hat{\cal S}, m \in {\bf Z}_{\ge 1} \}$
acting on the common quantum space as above.
To do so, we first recall the $T$-system for the
corresponding non-twisted cases
$X^{(1)}_n = A^{(1)}_n, D^{(1)}_n, E^{(1)}_6$ and
$D^{(1)}_4$ [2].
The $T$-system in these cases has the simple form
$(a \in {\cal S}, m \in {\bf Z}_{\ge 1})$
$$
T^{(a)}_m(u+1)T^{(a)}_m(u-1) = T^{(a)}_{m+1}(u)T^{(a)}_{m-1}(u)
+ g^{(a)}_m(u) \prod_{b \in {\cal S}_a}T^{(b)}_m(u).
\eqno(3.3)$$
Here ${\cal S}_a$ stands for the set of adjacent nodes
to $a \in {\cal S}$ on the $X_n$-Dynkin diagram.
The $g^{(a)}_m(u)$ is a
scalar function analogous to the
quantum determinant and depends on the quantum space choice.
It satisfies
$g^{(a)}_m(u+1)g^{(a)}_m(u-1) =
g^{(a)}_{m+1}(u)g^{(a)}_{m-1}(u)$ [2].
The $T$-system for $X^{(\kappa)}_n$ is formally obtained from (3.3)
by further imposing the
``modulo $\sigma$ relations":
$$\eqalignno{
T^{(a)}_m(u) &= \epsilon^a_m
T^{(\sigma(a))}_m(u+{\pi i \over \kappa\hbar}),
&(3.4{\rm a})\cr
(\epsilon^a_m )^\kappa &= (-)^{DN},&(3.4{\rm b})\cr}
$$
where $D$ will be explained after (4.3).
By means of (3.4a), one can confine
(3.3) into the equations only
among $T^{(a)}_m(u)$ with $a \in \hat{\cal S}$.
We call the the resulting functional relation the
$X^{(\kappa)}_n$ $T$-system.
Apart from the $a = \sigma(a)$ case in (3.4),
it reads as follows.
($T^{(a)}_0(u) = T^{(0)}_m(u) = 1$.)
$$\eqalign{A^{(2)}_{2r}:&\cr
&T^{(a)}_m(u+1)T^{(a)}_m(u-1) = T^{(a)}_{m+1}(u)T^{(a)}_{m-1}(u)\cr
&\qquad\qquad\qquad\qquad\qquad
+ g^{(a)}_m(u) T^{(a+1)}_m(u)T^{(a-1)}_m(u), 1 \le a \le r-1,\cr
&T^{(r)}_m(u+1)T^{(r)}_m(u-1) = T^{(r)}_{m+1}(u)T^{(r)}_{m-1}(u)\cr
&\qquad\qquad\qquad\qquad\qquad
+ g^{(r)}_m(u) T^{(r)}_m(u+{\pi i \over 2 \hbar})
T^{(r-1)}_m(u).\cr}\eqno(3.5)
$$
$$\eqalign{A^{(2)}_{2r-1}:&\cr
&T^{(a)}_m(u+1)T^{(a)}_m(u-1) = T^{(a)}_{m+1}(u)T^{(a)}_{m-1}(u)\cr
&\qquad\qquad\qquad\qquad\qquad
+ g^{(a)}_m(u) T^{(a+1)}_m(u)T^{(a-1)}_m(u), 1 \le a \le r-1,\cr
&T^{(r)}_m(u+1)T^{(r)}_m(u-1) = T^{(r)}_{m+1}(u)T^{(r)}_{m-1}(u)\cr
&\qquad\qquad\qquad\qquad\qquad
+ g^{(r)}_m(u) T^{(r-1)}_m(u+{\pi i \over 2 \hbar})
T^{(r-1)}_m(u).\cr}
\eqno(3.6)
$$
$$\eqalign{D^{(2)}_{r+1}:&\cr
&T^{(a)}_m(u+1)T^{(a)}_m(u-1) = T^{(a)}_{m+1}(u)T^{(a)}_{m-1}(u)\cr
&\qquad\qquad\qquad\qquad\qquad
+ g^{(a)}_m(u) T^{(a+1)}_m(u)T^{(a-1)}_m(u), 1 \le a \le r-2,\cr
&T^{(r-1)}_m(u+1)T^{(r-1)}_m(u-1) =
T^{(r-1)}_{m+1}(u)T^{(r-1)}_{m-1}(u)\cr
&\qquad\qquad\qquad\qquad\qquad
+ g^{(r-1)}_m(u) T^{(r-2)}_m(u)T^{(r)}_m(u)
T^{(r)}_m(u+{\pi i \over 2 \hbar}),\cr
&T^{(r)}_m(u+1)T^{(r)}_m(u-1) = T^{(r)}_{m+1}(u)T^{(r)}_{m-1}(u)
+ g^{(r)}_m(u) T^{(r-2)}_m(u).\cr}\eqno(3.7)
$$
$$\eqalign{E^{(2)}_6:&\cr
&T^{(1)}_m(u+1)T^{(1)}_m(u-1) = T^{(1)}_{m+1}(u)T^{(1)}_{m-1}(u)
+ g^{(1)}_m(u) T^{(2)}_m(u),\cr
&T^{(2)}_m(u+1)T^{(2)}_m(u-1) = T^{(2)}_{m+1}(u)T^{(2)}_{m-1}(u)
+ g^{(2)}_m(u) T^{(1)}_m(u)T^{(3)}_m(u),\cr
&T^{(3)}_m(u+1)T^{(3)}_m(u-1) = T^{(3)}_{m+1}(u)T^{(3)}_{m-1}(u)\cr
&\qquad\qquad\qquad\qquad\qquad
+ g^{(3)}_m(u) T^{(2)}_m(u)
T^{(2)}_m(u+{\pi i \over 2\hbar})T^{(4)}_m(u),\cr
&T^{(4)}_m(u+1)T^{(4)}_m(u-1) = T^{(4)}_{m+1}(u)T^{(4)}_{m-1}(u)
+ g^{(4)}_m(u) T^{(3)}_m(u).\cr}\eqno(3.8)
$$
$$\eqalign{D^{(3)}_4:&\cr
&T^{(1)}_m(u+1)T^{(1)}_m(u-1) = T^{(1)}_{m+1}(u)T^{(1)}_{m-1}(u)
+ g^{(1)}_m(u) T^{(2)}_m(u),\cr
&T^{(2)}_m(u+1)T^{(2)}_m(u-1) = T^{(2)}_{m+1}(u)T^{(2)}_{m-1}(u)\cr
&\qquad\qquad\qquad\qquad\qquad
+ g^{(2)}_m(u) T^{(1)}_m(u)
T^{(1)}_m(u+{\pi i \over 3\hbar})
T^{(1)}_m(u-{\pi i \over 3\hbar}).\cr}\eqno(3.9)
$$
In the above, we have absorped the $\epsilon^a_m$ factors
into redifinitions of $g^{(a)}_m(u)$.
The $A^{(2)}_2$ case of (3.5)
agrees with eq.(15) of [5] for the Izergin-Korepin model [21].
\par
Our proposal (3.5-9) has mainly stemmed from the analytic Bethe
ansatz study of the transfer matrix eigenvalues
similar to [12].
By using the $U_q(X^{(\kappa)}_n)$-Bethe equation
having a ``modulo $\sigma$ structure" [22] (see (4.2)),
one can consistently build the DVFs as in
[12] to convince oneself
that the eigenvalues $\Lambda^{(a)}_m(u)$ hence
$T^{(a)}_m(u)$ will obey (3.4).
This will be actually seen in sections 4, 5 and 6.
\par
If one dropps the $u$-dependence totally by setting
$T^{(a)}_m(u+\cdots) = \hat{Q}^{(a)}_m,
g^{(a)}_m(u) = 1$ and $\epsilon^a_m = 1$,
the resulting equations on $\hat{Q}^{(a)}_m$
have a solution
$$
\hat{Q}^{(a)}_m = Q^{(\dot{a})}_m \vert_{\hbox{specialization}},\quad
a \in \hat{\cal S}.
\eqno(3.10)
$$
Here $Q^{(\dot{a})}_m$ denotes the Yangian analogue of the
character that satisfies the
$Q$-system for $X_n$ [18,19,2].
The specialization on the RHS is done
so that the $\sigma$-invariance
$Q^{(b)}_m = Q^{(\sigma(b))}_m (\forall b \in {\cal S})$ is achieved.
Certainly (3.10) becomes (2.4) under such a
specialization where the characters are reduced to the dimensions.
\par
As in section 5 of [2],
one can express $T^{(a)}_m(u) \, (a \in \hat{\cal S})$ in terms of
$T^{(a)}_1(u+\hbox{shifts})$ by applying the $T$-system (3.5-9)
recursively.
The result is simply related to the non-twisted case [2]
via (3.4) with $\epsilon^a_m = 1$.
\par\vskip0.5cm\noindent
{\bf 4. Analytic Bethe ansatz}\par
For $a \in {\cal S}$, let $\alpha_a$ and  $\omega_a$ be the
simple root and the fundamental weight of
$X_n = A_n, D_n, E_6$, respectively.
We employ the standard normalization
$(\alpha_a \vert \alpha_a) = 2,
(\alpha_a \vert \omega_b) = \delta_{a b}$
via the bilinear form $(\,\,\vert \,\,)$.
Define the functions
$$\eqalignno{
[u]_k &= q^{ku} - q^{-ku},&(4.1{\rm a})\cr
\phi^k(u) &= \prod_{j=1}^N [u - w_j]_k,&(4.1{\rm b})\cr
Q^k_a(u) &= \prod_{j=1}^{N_a}[u - iu^{(a)}_j]_k\quad
\hbox{ for } a \in \hat{\cal S},&(4.1{\rm c})\cr}
$$
where $N_a$ is a non-negative integer.
When $k=1$, we simply write as
$\phi(u) = \phi^1(u)$ and $Q_a(u) = Q^1_a(u)$.
There should be no confusion between
$Q^k_a(u)$ here and $Q^{(\dot{a})}_m$ in (3.10).
In this notation, the $U_q(X^{(\kappa)}_n)$-Bethe ansatz equation
(BAE) [22] is given as follows.
$$
-\prod_{t=0}^{\kappa-1}
{\phi(iu^{(a)}_j + (s\omega_{\dot{p}}\vert\alpha_{\sigma^t(\dot{a})})+
{t\pi i\over \kappa\hbar})\over
\phi(iu^{(a)}_j - (s\omega_{\dot{p}}\vert\alpha_{\sigma^t(\dot{a})})+
{t\pi i\over \kappa\hbar})} =
\prod_{t=0}^{\kappa-1}\prod_{b\in \hat{\cal S}}
{Q_b(iu^{(a)}_j+(\alpha_{\dot{a}} \vert \alpha_{\sigma^t(\dot{b})})
+{t\pi i\over \kappa\hbar})\over
Q_b(iu^{(a)}_j-(\alpha_{\dot{a}}\vert \alpha_{\sigma^t(\dot{b})})
+{t\pi i\over \kappa\hbar})}.
\eqno(4.2)
$$
For any quantum space choice labelled by
$p \in \hat{\cal S}$ and $s \in {\bf Z}_{\ge 1}$,
this is a system of simultaneous equations
on the complex numbers
$\{ u^{(a)}_j \mid a \in \hat{\cal S}, 1 \le j \le N_a \}$.
The $Q^k_a(u)$ (4.1c) is defined for each solution to the above
BAE.
\par
In the rest of this section we briefly observe
a few features in the analytic Bethe ansatz [23]
based on (4.2).
This method is a hypothesis that
the transfer matrix eigenvalue
in expressed in the DVF:
$$\eqalignno{
\Lambda^{(a)}_m(u) &= \sum_{h=1}^{\dim \hat{W}^{(a)}_m} T_h,
&(4.3{\rm a})\cr
T_h &= (dr T_h)(vac T_h),&(4.3{\rm b})\cr
dr T_h &=
{Q^{k^h_1}_{a^h_1}(u+x^h_1) \cdots Q^{k^h_{j(h)}}_{a^h_{j(h)}}(u+x^h_{j(h)})
\over
Q^{k^h_1}_{a^h_1}(u+y^h_1) \cdots Q^{k^h_{j(h)}}_{a^h_{j(h)}}(u+y^h_{j(h)})},
&(4.3{\rm c})\cr
vac T_h &= \phi^{l^h_1}(u+z^h_1) \cdots \phi^{l^h_{d(h)}}(u+z^h_{d(h)}).
&(4.3{\rm d})\cr}
$$
Here (4.3c) and (4.3d) are
called the dress part and the vacuum part, respectively.
$d(h), j(h), k^h_i, l^h_i, a^h_i, x^h_i, y^h_i$ and $z^h_i$
are to be determined
so that (4.3) fulfills a couple of
conditions inherited from the properties of the relevant
$R$-matrix
$R_{\hat{W}^{(a)}_m,\hat{W}^{(p)}_s}(u)$ [23,12].
In particular, they are to be chosen so that
$\Lambda^{(a)}_m(u)$ becomes pole-free on condition that
the BAE (4.2) is valid.
The sum $D = l^h_1 + \cdots + l^h_{d(h)}$ is
independent of $h$.
Let
$$\eqalignno{
&\sum_{h=1}^{\dim W^{(\dot{a})}_m} \dot{T}_h,
&(4.4{\rm a})\cr
&\dot{T}_h = (dr \dot{T}_h)(vac \dot{T}_h)&(4.4{\rm b})\cr}
$$
be the DVF for the eigenvalue of the
the transfer matrix
$T^{(\dot{a})}_m(u)$ for the $X^{(1)}_n$ case.
This corresponds to the non-twisted counterpart
of (4.3) in the light of (2.4).
In (4.4a), the dress part $dr \dot{T}_h$ is a ratio
of the $X^{(1)}_n$ analogue of $Q^1_a(u)$ (4.1c), which
are determined from the solutions to the BAE (4.2) for the
non-twisted case
$\kappa = 1, {\cal S} = \hat{\cal S}$.
Let us
write it as $\dot{Q}_a(u)\, (a \in {\cal S})$.
Because of (2.4), the sums
(4.3a) and (4.4a) consists of the same number of terms.
Moreover one can relate the dress parts as
$$\eqalign{
dr T_h &= dr \dot{T}_h\vert_{\sigma -\hbox{reduction}},\cr
\sigma -\hbox{reduction}:& \,\,\dot{Q}_a(u) \rightarrow
\prod_{t=0}^{\kappa-1}
\bigl(Q_{\hat{a}}(u-{t\pi i\over \kappa\hbar})\bigr)^{\nu(a,t)},
\quad \hbox{for all } a \in {\cal S},\cr
\nu(a,t) &= \cases{1& if $\sigma^t(a) = \dot{\hat{a}}$\cr
                 0& otherwise}.\cr}\eqno(4.5)
$$
This follows simply by comparing (4.2)
with the non-twisted case $\kappa = 1$.
It has been already used implicitly in [20] and will actually be
seen in sections 5 and 6.
\par
Let us include a remark before closing this section.
Suppose one has found the DVF
when the quantum space is
$\otimes_{j=1}^N W^{(p)}_1(w_j)$.
Then, the one for
$\otimes_{j=1}^N W^{(p)}_s(w_j)$
can be deduced from it by the
replacement
$$
\phi^k(u) \rightarrow
\prod_{j=1}^s\phi^k(u+s+1-2j).
$$
See (4.1) and (4.2).
We shall
henceforth consider the $s=1$ case only with no loss of generality.
\par\vskip0.5cm
\noindent
{\bf 5. Solutions for $A^{(2)}_n$}\par
Here we give the solution of the $T$-system
(3.5,6) for $A^{(2)}_n = A^{(2)}_{2r}$ and $A^{(2)}_{2r-1}$.
Define a set $J$ by
$$
 J = \cases{  \{1,2,\cdots, r,\overline{r},\cdots,
                 \overline{2},\overline{1} \}
              & for $A^{(2)}_{2r-1}$, \cr
               \{1,2,\cdots r,0,\overline{r},\cdots,
                 \overline{2},\overline{1} \}
              & for $A^{(2)}_{2r}$, \cr
              }\eqno(5.1)
$$
and specify the ordering among these letters as
$$\eqalign{
&1 \prec 2 \prec \cdots   \prec r
 \prec \overline{r} \prec \cdots \prec
 \overline{2} \prec \overline{1} \quad \hbox{ for } A^{(2)}_{2r-1},\cr
&1 \prec 2 \prec \cdots   \prec r \prec 0
 \prec \overline{r} \prec \cdots \prec
 \overline{2} \prec \overline{1}
\quad \hbox{ for } A^{(2)}_{2r}.\cr}\eqno(5.2)
$$
We introduce the boxes that contain an element of
$J$ and represent a dressed vacuum as follows.
$$\eqalign{
\Fsquare(0.5cm,a)  &= \psi_{a}(u)
      {{Q_{a-1}( u+a+1 )
            Q_{a}(u+a-2 )}\over
       { Q_{a-1}(u+a-1)Q_{a}(u+a)}}
  \quad 1 \le a \le [{n\over 2}],\cr
\Fsquare(0.5cm,\bar{a})  &=
\psi_{\bar{a}}(u)
      {{Q_{a-1}( u+n-a+{\pi i\over 2\hbar}  )
        Q_{a}(u+n-a+3+{\pi i\over 2\hbar} )}\over
       { Q_{a-1}(u+n-a+2+{\pi i\over 2\hbar} )
         Q_{a}(u+n-a+1+{\pi i\over 2\hbar} )}}
  \quad 1 \le a \le [{n\over 2}],\cr
\Fsquare(0.5cm,r)  &= \psi_{ r}(u)
    {{ Q_{r-1}(u+r+1) Q^2_{r}(u+r-2)}
            \over {Q_{r-1}(u+r-1)Q^2_{r}(u+r)} }
  \quad \hbox{ for } A^{(2)}_{2r-1},\cr
\Fsquare(0.5cm,\overline{r})  &=  \psi_{ \overline{r} }(u)
    {{ Q_{r-1}(u+r-1+{\pi i\over 2\hbar})
       Q^2_{r}(u+r+2)}
\over {Q_{r-1}(u+r+1+{\pi i\over 2\hbar})Q^2_{r}(u+r)} }
  \quad \hbox{ for } A^{(2)}_{2r-1},        \cr
\Fsquare(0.5cm, 0)  &=
  \psi_0(u) {{Q_r(u+r+2) Q_{r}(u+r-1+{\pi i\over 2\hbar})}\over
   { Q_{r}(u+r)Q_{r}(u+r+1+{\pi i\over 2\hbar})}}
  \quad \hbox{ for } A^{(2)}_{2r}, \cr}\eqno(5.3{\rm a})
$$
where we have put $Q_0(u)=1$.
The symbol $[x]$ stands for the largest integer
not exceeding $x$.
In the above, the spectral parameter $u$ is attached
to each box.
We shall exhibit the $u$-dependence as
$\Fsquare(0.5cm, a)_u$ when necessary.
The functions $\psi_a(u)$ are the
vacuum parts and depend on the choice
of the quantum space.
For $s=1, 1\le p \le r$, they are given by
$$\eqalign{
\psi_{1}(u)&=\cdots =\psi_{p}(u)=\phi(u+p+1)
\phi(u+n+2-p+{\pi i\over 2\hbar}), \cr
\psi_{p+1}(u)&=\cdots =
     \psi_{\overline{p+1}}(u)=\phi(u+p-1)
\phi(u+n+2-p+{\pi i\over 2\hbar}), \cr
\psi_{\overline{p}}(u)&=\cdots =
     \psi_{\overline{1}}(u)=\phi(u+p-1)\phi(u+n-p+{\pi i\over 2\hbar}).
}\eqno(5.3{\rm b})
$$
Note that the second possibility is void for
$A^{(2)}_{2r-1}$ and $p=r$.
\par
Now we introduce the set ${\cal T}^{(a)}_m$ of the
semi-standard tableaux with
$a \times m$ rectangular shape;
$$
{\cal T}^{(a)}_m =
\Bigl \{
	\normalbaselines\m@th\offinterlineskip
	\vcenter{
   \hbox{$\Flect(0.6cm,0.6cm,\hbox{$i_{1 1}$})\hskip-0.4pt
         \Flect(0.6cm,1.5cm,\hbox{$\cdots$}) \hskip-0.4pt
         \Flect(0.6cm,0.6cm,\hbox{$i_{1 m}$})$}
    \vskip-0.4pt
   \hbox{$\Flect(1.2cm,0.6cm,\hbox{$\vdots$})\hskip-0.4pt
         \Flect(1.2cm,1.5cm,\hbox{$\ddots$}) \hskip-0.4pt
         \Flect(1.2cm,0.6cm,\hbox{$\vdots$})$
         }
     \vskip-0.4pt
   \hbox{$\Flect(0.6cm,0.6cm,\hbox{$i_{a 1}$})\hskip-0.4pt
         \Flect(0.6cm,1.5cm,\hbox{$\cdots$}) \hskip-0.4pt
         \Flect(0.6cm,0.6cm,\hbox{$i_{a m}$})$}
         }
 \quad \vrule \quad i_{j k} \in J, \,
i_{1 k}\prec \cdots \prec i_{a k}, \,
i_{j 1} \preceq \cdots \preceq i_{j m}
\Bigr \}.\eqno(5.4)
$$
We identify each element of ${\cal T}^{(a)}_m$ as above with
the product
$$
\prod_{j=1}^a \prod_{k=1}^m
\Fsquare(0.6cm, i_{j k})_{u+a-m-2j+2k}.
\eqno(5.5)
$$
Put
$$\eqalignno{
\Lambda^{(a)}_m(u) &= {1 \over {f^{(a)}_m(u)}}
 \sum_{ T \in {\cal T}^{(a)}_m} T, &(5.6{\rm a})\cr
f^{(a)}_m(u) &=\prod_{j=1}^{m} f^{(a)}_1(u+m+1-2j),
 &(5.6{\rm b})  \cr
f^{(a)}_1(u) &=
\bigl(\prod_{j=1}^{\vert a-p\vert} \phi(u+\vert a-p\vert-2j)
     \phi(u-\vert a-p\vert+n+1+2j+{\pi i\over 2\hbar})
\bigr)^{\pm 1},&(5.6{\rm c})\cr}
$$
where the power in (5.6c) should be chosen as
$+1$ or $-1$ according to $a \ge p$ or $a < p$, respectively.
One can check that
for $a > p$ the denominator $f^{(a)}_m(u)$ in (5.6a)
can be completely
cancelled out for any $T \in {\cal T}^{(a)}_m$.
Consequently, the vacuum parts in $\Lambda^{(a)}_m(u)$
are homogeneous of order $mp$ with respect to
both $\phi(u+\xi)$ and $\phi(u+\eta+{\pi i\over 2\hbar})$,
where $\xi, \eta \in {\bf Z}$.
Now the main claim in this section is
\proclaim Theorem.
For $1 \le p \le r$, the above $\Lambda^{(a)}_m(u)$
is pole-free provided that the BAE (4.2) is valid.
It satisfies the
$A^{(2)}_n$ $T$-system (3.5,6) with
$$\eqalign{
    g^{(a)}_m(u)&=1 \qquad \hbox{ for }  1 \le a \le r-1,  \cr
    g^{(r)}_m(u)&=(-)^{Nmp}.   \cr
}\eqno(5.7)$$
\par
Due to (4.5), the proof essentially reduces to that for $A^{(1)}_n$ case,
which can be done by combining the results in [7] and [2].
Based on the analytic Bethe ansatz,
we thus suppose that (5.6) gives the eigenvalues
of the fusion $A^{(2)}_n$ vertex models.
For $p=1$, $\Lambda^{(1)}_1(u)$ actually coincides with the DVF
obtained earlier in [20].
For $A^{(2)}_{2r-1}$, the property
$\Lambda^{(r)}_m(u) = (-)^{Nmp}\Lambda^{(r)}_m(u+{\pi i\over 2\hbar})$
holds, which corresponds to (3.4) for the fixed node
$r = \sigma(r)$.
\par\vskip0.5cm\noindent
{\bf 6. Conjectures for $D^{(3)}_4$}\par
A peculiarity of $D^{(3)}_4$ is
that this is the only twisted affine Lie
algebra with $\kappa = 3$.
In this section, we will give conjectures on
the solutions to the $T$-system (3.9) for this algebra.
First, we define a set $J$
$$
 J=\{1,2,3,4,\overline{4},\overline{3},
\overline{2},\overline{1} \}\eqno(6.1)
$$
and specify an order in it as
$$
1 \prec 2 \prec 3   \prec
\eqalign{&4\cr &\overline{4}\cr} \prec \overline{3} \prec
 \overline{2} \prec \overline{1}. \eqno(6.2)
$$
We assume no order between $4$ and $\overline{4}$.
These letters have an origin in the labels for $D_4$ [12].
As in the $A^{(2)}_n$ case, we introduce
elementary boxes carrying $a \in J$ and the
spectral parameter $u$ by
$$\eqalign{
\Fsquare(0.5cm,1)  &= \psi_{1}(u)
      {{Q_{1}( u-1  ) }\over{ Q_{1}(u+1)}},  \cr
\Fsquare(0.5cm,2)  &= \psi_{2}(u)
      {{Q_{1}(u+3)Q_{2}^{3}(u) }\over
       { Q_{1}(u+1) Q_{2}^{3}(u+2)}}  \cr
\Fsquare(0.5cm,3)  &= \psi_{3}(u)
      {{Q_{1}(u+1+{\pi i\over 3\hbar})
        Q_{1}(u+1-{\pi i\over 3\hbar}) Q_{2}^{3}(u+4) }\over
       { Q_{1}(u+3+{\pi i\over 3\hbar})
        Q_{1}(u+3-{\pi i\over 3\hbar})
       Q_{2}^{3}(u+2)}},  \cr
\Fsquare(0.5cm,4)  &= \psi_{4}(u)
      {{Q_{1}(u+1-{\pi i\over 3\hbar})
        Q_{1}(u+5+{\pi i\over 3\hbar}) }\over
       { Q_{1}(u+3-{\pi i\over 3\hbar})
         Q_{1}(u+3+{\pi i\over 3\hbar}) }},  \cr
\Fsquare(0.5cm,\overline{4})  &= \psi_{\overline{4}}(u)
      {{Q_{1}(u+1+{\pi i\over 3\hbar})
        Q_{1}(u+5-{\pi i\over 3\hbar}) }\over
       { Q_{1}(u+3+{\pi i\over 3\hbar})
         Q_{1}(u+3-{\pi i\over 3\hbar}) }},  \cr
\Fsquare(0.5cm,\overline{3})  &= \psi_{\overline{3}}(u)
      {{Q_{1}(u+5+{\pi i\over 3\hbar})
        Q_{1}(u+5-{\pi i\over 3\hbar})
        Q_{2}^{3}(u+2) }\over
       { Q_{1}(u+3+{\pi i\over 3\hbar})
         Q_{1}(u+3-{\pi i\over 3\hbar})
         Q_{2}^{3}(u+4)}},  \cr
\Fsquare(0.5cm,\overline{2})  &= \psi_{\overline{2}}(u)
      {{Q_{1}(u+3)Q_{2}^{3}(u+6) }\over
       { Q_{1}(u+5) Q_{2}^{3}(u+4)}},  \cr
\Fsquare(0.5cm,\overline{1})  &= \psi_{\overline{1}}(u)
      {{Q_{1}( u+7) }\over{ Q_{1}(u+5)}}.  \cr
}\eqno(6.3{\rm a})$$
In the above, $\psi_{a}(u)$ denotes the
vacuum part and depends on the quantum space choice.
For $s=1$ and $p=1,2$ they read
$$
\eqalign{
p=1:&  \cr
&\psi_1(u)=\phi(u+2)\phi(u+6)\phi(u+4+{\pi i\over 3\hbar})
\phi(u+4-{\pi i\over 3\hbar}),  \cr
&\psi_2(u)=\psi_3(u)=
   \phi(u)\phi(u+6)\phi(u+4+{\pi i\over 3\hbar})
\phi(u+4-{\pi i\over 3\hbar}),  \cr
&\psi_4(u)=\phi(u)\phi(u+6)\phi(u+2+{\pi i\over 3\hbar})
\phi(u+4-{\pi i\over 3\hbar}),  \cr
&\psi_{\overline{4}}(u)=
          \phi(u)\phi(u+6)\phi(u+4+{\pi i\over 3\hbar})
\phi(u+2-{\pi i\over 3\hbar}),  \cr
&\psi_{\overline{3}}(u)=\psi_{\overline{2}}(u)=
          \phi(u)\phi(u+6)\phi(u+2+{\pi i\over 3\hbar})
\phi(u+2-{\pi i\over 3\hbar}),  \cr
&\psi_{\overline{1}}(u)=
          \phi(u)\phi(u+4)\phi(u+2+{\pi i\over 3\hbar})
\phi(u+2-{\pi i\over 3\hbar}), \cr
p=2:&  \cr
&\psi_1(u)=\psi_2(u)=
          \phi^3(u+3) \phi^3(u+5),   \cr
&\psi_3(u)=\psi_{\overline{3}}(u)=
\psi_4(u)=\psi_{\overline{4}}(u)=
          \phi^3(u+1) \phi^3(u+5),   \cr
&\psi_{\overline{2}}(u)=\psi_{\overline{1}}(u)=
          \phi^3(u+1) \phi^3(u+3).
}\eqno(6.3{\rm b})$$
Next we consider the following sets
${\cal T}^{(1)}_m$ and ${\cal T}^{(2)}_m$ of the
tableaux containing the elements of $J$.
$$\eqalign{
{\cal T}^{(1)}_m &=  \Bigl \{
\fsquare(0.4cm,i_1)\naga\fsquare(0.4cm,i_m)
 \, \, \vrule  \quad
  i_k \preceq i_{k+1}  \hbox{ and }
               (i_k,i_{k+1}) \not= (4,\overline{4}),
                            (\overline{4}, 4)
                \hbox{ for } 1 \le k <m  \Bigr \},  \cr
{\cal T}^{(2)}_m &=  \Bigl \{
	\normalbaselines\m@th\offinterlineskip
	\vcenter{
   \hbox{$\Flect(0.6cm,0.6cm,\hbox{$i_{1 1}$})\hskip-0.4pt
         \Flect(0.6cm,1.5cm,\hbox{$\cdots$}) \hskip-0.4pt
         \Flect(0.6cm,0.6cm,\hbox{$i_{1 m}$})$}
    \vskip-0.4pt
   \hbox{$\Flect(0.6cm,0.6cm,\hbox{$i_{2 1}$})\hskip-0.4pt
         \Flect(0.6cm,1.5cm,\hbox{$\cdots$}) \hskip-0.4pt
         \Flect(0.6cm,0.6cm,\hbox{$i_{2 m}$})$}
            }
       \,\,  \vrule  \quad
 \hbox{conditions} \, \hbox{(i)} - \hbox{(iv)}
 \, \hbox{below}\Bigr \},
}$$
$$\eqalign{
\hbox{(i)}&\,\,\hbox{both rows belong to } {\cal T}^{(1)}_m,\cr
\hbox{(ii)}&\,\,i_{1 k} \prec i_{2 k} \hbox{ or }
(i_{1 k},i_{2 k})=(4,\overline{4}), (\overline{4}, 4)
\hbox{  for  } 1 \le k \le m,\cr
\hbox{(iii)}&\,\,\hbox{the columns }
	\normalbaselines\m@th\offinterlineskip
	\vcenter{
   \hbox{\Fsquare(0.5cm,3)}
	      \vskip-0.4pt
   \hbox{\Fsquare(0.5cm,4)} }
 \hbox{ and }
	\normalbaselines\m@th\offinterlineskip
	\vcenter{
   \hbox{\Fsquare(0.5cm,4)}
	      \vskip-0.4pt
   \hbox{\Fsquare(0.5cm,\overline{3})} }
\hbox{  do not appear simultaneously, }\cr
\hbox{(iv)}&\,\,\hbox{the columns }
	\normalbaselines\m@th\offinterlineskip
	\vcenter{
   \hbox{\Fsquare(0.5cm,3)}
	      \vskip-0.4pt
   \hbox{\Fsquare(0.5cm,\overline{4})} }
 \hbox{ and }
	\normalbaselines\m@th\offinterlineskip
	\vcenter{
   \hbox{\Fsquare(0.5cm,\overline{4})}
	      \vskip-0.4pt
   \hbox{\Fsquare(0.5cm,\overline{3})} }
\hbox{  do not appear simultaneously. }\cr
}\eqno(6.4)$$
We identify elements in ${\cal T}^{(1)}_m$ and
${\cal T}^{(2)}_m$ with the dressed vacuums
by the same rule as (5.5) but with the elementary
boxes defined by (6.3).
\par
For $a = 1,2$ we now put
$$\eqalignno{
\Lambda^{(a)}_m(u) &=
  \Bigl({1\over {g^{(1)}_m(u)}}\Bigr)^{\delta_{a 2}}
  \sum_{T \in {\cal T}^{(a)}_m} T,
&(6.5{\rm a})\cr
g^{(a)}_m(u) &=\prod_{j=1}^m g^{(a)}_1(u+m+1-2j),
&(6.5{\rm b})\cr
g^{(a)}_1(u) &=
\cases{ \phi^{2a-1}(u-1) \phi^{2a-1}(u+7)
& if $a=p$ \cr
1 & otherwise \cr}.  &(6.5{\rm c})\cr}
$$
One can show that the denominator $g^{(1)}_m(u)$
in (6.5a) can be completely cancelled out for any
$T \in {\cal T}^{(2)}_m$.
Based on an extensive computer check, we have
\proclaim Conjecture. The $\Lambda^{(a)}_m(u)$ (6.5) is pole-free
provided that the BAE (4.2) is valid.
It satisfies the $D^{(3)}_4$ $T$-system (3.9).
\par
A few remarks are in order.
First, the DVF for $\Lambda^{(1)}_1(u)$ with
$p=1$ agrees with the one in [20], which is actually pole-free.
Second, the pole freeness can be checked directly for
$\Lambda^{(a)}_1(u), a, p \in \{1,2 \}$ from the explicit form (6.5).
Third, the property
$$
\Lambda^{(2)}_1(u+{\pi i \over 3\hbar}) =\Lambda^{(2)}_1(u)
$$
holds, which is consistent with
(3.4a) for the fixed node $2 = \sigma(2)$.
Finally, one can
set up a similar conjecture to the above also for the $D^{(1)}_4$
$T$-system [2] by using (4.5) and the same sets
${\cal T}^{(a)}_m$ in (6.4).
\par\vskip0.5cm\noindent
{\bf 7. Summary and discussion}\par
In this paper we have
proposed the $T$-system, the transfer matrix functional relations, for
the fusion vertex models associated with
the twisted quantum affine algebras
$U_q(X^{(\kappa)}_n)$ for
$X^{(\kappa)}_n = A^{(2)}_n, D^{(2)}_n,
E^{(2)}_6$ and $D^{(3)}_4$.
It extends the non-twisted case in [2,9].
We have also constructed the DVFs
that satisfy or conjecturally satisfy the
$A^{(2)}_n$ or $D^{(3)}_4$ $T$-system, respectively.
This is a yet further result along the scheme of [12].
Those DVFs have been derived essentially from the
non-twisted cases by noting the $\sigma$-reduction
relation (4.5).
Thus a similar construction of the DVF will also be possible
for the $D^{(2)}_n$ case based on the $D^{(1)}_n$ result in [12].
In view of
the analytic Bethe ansatz,
these DVFs are the candidates of the transfer matrix
eigenvalues for the fusion vertex models associated with
$U_q(X^{(\kappa)}_n)$.
The $T$-system here will work
efficiently for computing the
physical quantities as in [9].
We leave it for a future study.
\par
There are some issues that deserve further investigations.
The $T$-system proposed in this paper
applies to the vertex models, which is the {\it unrestricted} version
in the sense of [2].
It will be interesting to introduce
the level parameter $\ell$ and seek the {\it restricted} $T$-system
that applies to the level $\ell$ RSOS models as done in [2].
As for this, the level-rank duality between
the level $\ell$
$A^{(2)}_{2r-1}$ RSOS model and level $r$ $C^{(1)}_\ell$ RSOS
model [16] will be a good guide.
\par
One may set, for example in $D^{(3)}_4$ case, as
$$\eqalign{
y^{(1)}_m(u) &= {g^{(1)}_m(u)\Lambda^{(2)}_m(u)
\over  \Lambda^{(1)}_{m+1}(u)\Lambda^{(1)}_{m-1}(u)},\cr
y^{(2)}_m(u) &= {g^{(2)}_m(u)\Lambda^{(1)}_m(u)
\Lambda^{(1)}_m(u+{\pi i\over 3\hbar})\Lambda^{(1)}_m(u-{\pi i\over 3\hbar})
\over  \Lambda^{(2)}_{m+1}(u)\Lambda^{(2)}_{m-1}(u)}.\cr
}
\eqno(7.1)
$$
Then the $T$-system (3.9) for the eigenvalues is transformed into
$$\eqalign{
y^{(1)}_m(u+1)y^{(1)}_m(u-1) &=
{1+y^{(2)}_m(u) \over
(1+y^{(1)}_{m+1}(u)^{-1})(1+y^{(1)}_{m-1}(u)^{-1})},\cr
y^{(2)}_m(u+1)y^{(2)}_m(u-1) &=
{(1+y^{(1)}_m(u))(1+y^{(1)}_m(u+{\pi i\over 3\hbar}))
(1+y^{(1)}_m(u-{\pi i\over 3\hbar})) \over
(1+y^{(2)}_{m+1}(u)^{-1})(1+y^{(2)}_{m-1}(u)^{-1})}.\cr
}
\eqno(7.2)
$$
which is independent of $g^{(a)}_m(u)$.
Equations like (7.2) were called the $Y$-{\it system} in [2,9].
It can be derived similarly for the other cases
$A^{(2)}_n, D^{(2)}_n$ and $E^{(2)}_6$.
For the non-twisted case, the $Y$-system
for general $X^{(1)}_r$ was extracted in [24]
from the thermodynamic Bethe ansatz (TBA) equation in [19].
Its structure reflects the employed string hypothesis [19].
It will be interesting to examine if
the $Y$-system obtained here indicates a proper way of
setting up the
string hypothesis and doing TBA for
$U_q(X^{(\kappa)}_n)$ symmetry models.
\vfill\eject
\beginsection References

\item{[1]}{R.J.Baxter, {\it Exactly solved models in statistical mechanics},
(Academic Press, London, 1982)}
\item{[2]}{A.Kuniba, T.Nakanishi and J.Suzuki, ``Functional relations
in solvable lattice models I: Functional relations and representation theory",
HUTP-93/A022, hep-th.9309137, Int. J. Mod. Phys. A in press}
\item{[3]}{P.P.Kulish and E.K.Sklyanin, in
{\it Lecture Notes in Physics} {\bf 151} (Springer, Berlin 1982)}
\item{[4]}{R.J.Baxter and
P.A.Pearce, J.Phys. A: Math. Gen.  {\bf 15} (1982) 897}
\item{[5]}{N.Yu.Reshetikhin, Lett.Math.Phys. {\bf 7} (1983) 205}
\item{[6]}{A.N.Kirillov and N.Yu.Reshetikhin, J.Phys.A: Math. Gen.
{\bf 20)} (1987) 1587}
\item{[7]}{V.V.Bazhanov and N.Yu.Reshetikhin,
J. Phys. A: Math. Gen. {\bf 23} (1990) 1477}
\item{[8]}{A.Kl\"umper and P.A.Pearce, Physica {\bf A183} (1992) 304}
\item{[9]}{A.Kuniba, T.Nakanishi and J.Suzuki, ``Functional relations
in solvable lattice models II: Applications",
HUTP-93/A023, hep-th.9310060, Int. J. Mod. Phys. A in press}
\item{[10]}{A. Kuniba, J. Phys. A: Math. Gen. {\bf 27} (1994) L113}
\item{[11]}{J. Suzuki, ``Fusion $U_q(G^{(1)}_2)$ vertex models and
analytic Bethe ans\" atze", hep-th.9405201, preprint}
\item{[12]}{A. Kuniba and J. Suzuki, ``Analytic Bethe ansatz for
fundamental representations of Yangians", hep-th.9406180, preprint}
\item{[13]}{V.G.Kac, {\it Infinite dimensional Lie algebras, 3rd ed.},
(Cambridge University Press, Cambridge, 1990)}
\item{[14]}{V.G.Drinfel'd, in {\it Proceedings of the ICM, Berkeley}
(AMS, Providence, 1987), }
\item{}{M.Jimbo, Lett. Math. Phys. {\bf 10} (1985) 63}
\item{[15]}{V.V.Bazhanov, Phys. Lett. {\bf B159} (1985) 321,}
\item{}{M.Jimbo, Commun. Math. Phys. {\bf 102} (1986) 537}
\item{[16]}{A. Kuniba, Nucl. Phys. {\bf B355} (1991) 801}
\item{[17]}{P.P.Kulish, N.Yu.Reshetikhin and E.K.Sklyanin,
Lett. Math. Phys. {\bf 5} (1981) 393}
\item{[18]}{A.N.Kirillov and N.Yu.Reshetikhin, J. Sov. Math. {\bf 52}
(1990) 3156}
\item{[19]}{A. Kuniba, Nucl. Phys. {\bf B389} (1993) 209}
\item{[20]}{N.Yu.Reshetikhin, Lett. Math. Phys. {\bf 14} (1987) 235}
\item{[21]}{A.G.Izergin and V.E.Korepin, Commun. Math. Phys.
{\bf 79} (1981) 303}
\item{[22]}{N.Yu.Reshetikhin and P.Wiegmann, Phys. Lett.
{\bf B189} (1987) 125}
\item{[23]}{N.Yu.Reshetikhin, Sov. Phys. JETP {\bf 57} (1983) 691}
\item{[24]}{A.Kuniba and T.Nakanishi, Mod. Phys. Lett. {\bf A7} (1992) 3487}

\bye